\documentclass[prl,superscriptaddress,twocolumn,preprintnumbers]{revtex4-2}

\usepackage{graphicx}% Include figure files
\usepackage{dcolumn}% Align table columns on decimal point
\usepackage{bm}% bold math
\begin{document}

\preprint{SI-HEP-2021-08}
\preprint{SFB-257-P3H-21-012}      

\title{The Heavy Quark Expansion for the Charm Quark}% Force line breaks with \\

\author{Th.~Mannel}
%\email{mannel@physik.uni-siegen.de}
\author{D.~Moreno}
%\email{Daniel.Moreno@uni-siegen.de}
\author{A.~A.~Pivovarov}
% \email{pivovarov@physik.uni-siegen.de}
\affiliation{Center for Particle Physics Siegen, Theoretische Physik 1, Universit\"at Siegen\\ 57068 Siegen, Germany}%

\date{\today}% It is always \today, today,

\begin{abstract}
		\noindent	We show that one can re-arrange the Heavy Quark Expansion for 
			inclusive weak decays of charmed hadrons in such a way that the 
			resulting expansion is an expansion in $\Lambda_{\rm QCD} / m_c$ 
			and $\alpha_s (m_c)$ with order-one coefficients. Unlike in the case
			of the bottom quark, the leading term includes not only the contribution 
			of the free-quark decay, but also a tower of terms related to matrix 
			elements of four quark operators. 
\end{abstract}


\maketitle

%\tableofcontents

\noindent The rigorous formulation of a systematic expansion of 
observables for hadrons containing a heavy quark $Q$ in powers of 
$\Lambda_{\rm QCD} / m_Q$~\cite{Shifman:1987rj,Eichten:1989zv,Isgur:1989vq,Grinstein:1990mj}
(for a textbook presentation and further references see~\cite{Manohar:2000dt}), 
in addition to the perturbative expansion in $\alpha_s(m_Q)$, has greatly extended 
the applicability of QCD to heavy hadron phenomenology. In particular, its application 
to inclusive processes in the form of the Heavy Quark Expansion 
(HQE)~\cite{Chay:1990da,Bigi:1992su,Manohar:1993qn}
i.e.\ the expansion of observables in powers of $\Lambda_{\rm QCD} / m_Q$ and $\alpha_s (m_Q)$
has become a standard tool to analyze inclusive heavy hadron decays, which has subsequently
been refined over the last three decades, including the calculation of perturbative as well as 
$\Lambda_{\rm QCD} / m_Q$ corrections to higher orders.   

While the HQE for bottom hadrons turns out to converge quite well, yielding 
precise predictions with controllable uncertainties (e.g. for
inclusive 
semileptonic 
decays~\cite{Benson:2003kp,Gambino:2013rza}), 
it seems to be far less useful in the case of charmed
hadrons. This is on the one hand due to the fact that $\Lambda_{\rm QCD} / m_c$ and $\alpha_s (m_c)$ 
are not particularly small. On the other hand, the HQE has contributions with enhanced coefficients which in 
the case of charm cause a breakdown of the HQE, leaving us only with the possibility to make qualitative statements. 

However, for some charm observables the techniques of HQE are quite accurate at the quantitative level, 
yielding predictions  with the precision of a few percent. Examples for such observables are the 
predictions for hadron masses and the exclusive semileptonic processes
$B \to D^{(*)} \ell \nu$, both based on a $1/m_c$ expansion. 
We take this as a hint, that it might be possible to define a ``charm Heavy Quark Expansion'' (cHQE)
which allows us to make more accurate predictions with controllable theoretical uncertainties. 

Looking at the lifetimes of weakly-decaying charmed hadrons it is obvious that the standard HQE is 
in trouble: The lifetimes of the ground-state charmed particles differ
by factors of two~\cite{Zyla:2020zbs} and even more when including the ground state
baryons. This can hardly be explained by a contribution suppressed 
by $(\Lambda_{\rm QCD} / m_c)^n$, where in the standard HQE one finds $n=3$. This is in sharp contrast 
to the case of bottom hadrons, where the relative lifetime differences are at the level of ten percent 
or less~\cite{Zyla:2020zbs}. 
 
Already almost three decades ago it has been noticed that the HQE contains special contributions which 
lead to enhanced coefficients in the HQE~\cite{Guberina:1979xw}. Within the HQE, these contributions 
are related to matrix elements of four quark operators which have Wilson coefficients that are enhanced 
by a phase-space factor $16 \pi^2$ relative to the leading 
term~\cite{Neubert:1996we,Lenz:2013aua}. Coefficients with this enhancement 
appear first at order $(\Lambda_{\rm QCD} / m_c)^3$ in the HQE, and for the charm quark the 
phase space enhancement 
can overwhelm the smallness of the factor $(\Lambda_{\rm QCD} / m_c)^3$. 

The successful applications of the HQE to charm are all related to observables where the matrix elements
of these four quark operators are suppressed. 
On the one hand, these can be suppressions by factors of 
$(\alpha_s (m_c)/4\pi)^2$. On the other hand, the matrix elements of the four quark operators are pretty 
well described by the ``Vacuum Saturation Ansatz'' 
(VSA, originally formulated in the context of nonleptonic $B$ decays in~\cite{Bauer:1986bm}), 
which leads to a strong suppression of the dangerous four quark contributions in inclusive 
semileptonic decays.    

In nonleptonic weak decays such enhanced contributions appear already at the
leading order in  $\alpha_s (m_c)$ and jeopardize the standard HQE for charmed hadrons. In the present
letter we show that these terms can be readily identified, and the HQE can be reorganized. 

Unlike in the case of bottom hadrons, the leading term in this cHQE is not just the decay 
of the ``free'' quark, it rather needs to be supplemented by a term which contains a matrix 
element of a nonlocal four quark operator. This contribution is sensitive to the flavour of the 
spectator quark and generates lifetime differences already at leading order. Physically this indicates 
that the presence of the spectator quark is essential for charm phenomenology. 

In the following we sketch how to set up cHQE, which systematically treats these enhanced contributions. 
The remaining expansion is an expansion in powers of $\Lambda_{\rm QCD} / m_c$ and $\alpha_s (m_c)$, 
however, now with coefficients of order unity. 

We start from the standard HQE for the decay rate, which is set up using the optical theorem by writing 
\begin{equation} \label{width}
	\Gamma \sim {\rm Im} \, \int d^4 x \, \langle D (v) | T \left[ H_{\rm eff} (x) H_{\rm eff} (0) \right] 
	| D(v) \rangle \,,
\end{equation}
and a subsequent operator product expansion (OPE) which yields an expansion in inverse powers of the
charm-quark mass  
\begin{equation} \label{OPE}
	\int d^4 x \,  T \left[ H_{\rm eff} (x) H_{\rm eff} (0) \right]  = \sum_{d=3}^\infty 
	\left(\frac{1}{m_c}\right)^{d-3}  \!\! \sum_i C_i^{(d)} O_i^{(d)}  \,,
\end{equation}
where $d$ is the dimension of the local operators $O_i^{(d)}$ and the index $i$ counts the 
operators of dimension $d$. The coefficients $C_i^{(d)}$ 
can be 
computed perturbatively as a power series in $\alpha_s (m_c)$. 

We consider first semileptonic charm decays, for which the effective Hamiltonian is 
given by 
\begin{equation}
	H_{\rm eff,sl}	= \frac{4 G_F}{\sqrt{2}} \sum_{q=d,s} V_{cq} (\bar{c}_L \gamma_\mu q_L) \,  
	(\bar{\ell}_L \gamma^\mu  \nu_{\ell\,L} )  \,. 
\end{equation}
Inserting this into Eq.~(\ref{width}) and taking the forward matrix element with $D$ meson states, 
we obtain 
\begin{eqnarray}  
	&& \Gamma^{sl} = \frac{G_F^2}{M_D}\sum_{q=d,s} |V_{cq}|^2 \int \frac{d^4q}{(2\pi)^4} 
	   \frac{1}{3 \pi} (q^\mu q^\nu - g^{\mu \nu} q^2)   
        \\
        &&  
        \times
	 \int d^4 x \, e^{-iqx} 
	\nonumber
	 \langle D (v) |  (\bar{c}_L \gamma_\mu q_L)(x)  \, (\bar{q}_L \gamma_\nu c_L)(0)   | D(v) \rangle \,,
\end{eqnarray} 
where we have contracted the leptonic part already.   

The leading terms of the OPE appear at $d=3$, and the coefficients $C_i^{(3)}$ are 
computed at tree level by contracting the light-quark line, leaving us with a two-quark 
operator of the form $\bar{c} \Gamma c$. Including the leptons, 
these coefficients are given in terms of the three-particle phase space of the 
(partonic) state consisting of $\ell$, $\bar{\nu}$ and $q$. This is also true for the 
coefficients up to (and including) order $1/m_c^2$, while starting at order $1/m_c^3$ 
one obtains a set of Wilson coefficients where the leading term is determined from a 
two-particle phase space, implying a phase space enhancement of a factor of $ 16 \pi^2$.   

These contributions are related to four quark operators, which in the semileptonic 
case read 
\begin{eqnarray} \label{4q} 
	&& \Gamma_{4q}^{sl} = \frac{G_F^2}{M_D} \sum_{q=d,s} |V_{cq}|^2 \int \frac{d^4q}{(2\pi)^4} 
	     \frac{1}{3 \pi} (q^\mu q^\nu - g^{\mu \nu} q^2)   	
	\\
	&&   
	\times
	  \int d^4 x \, e^{-iqx}
	\langle D (v) |  : (\bar{c}_L \gamma_\mu q_L)(x)  \,  (\bar{q}_L \gamma_\nu c_L)(0) : | D(v) \rangle \,,
	\nonumber
\end{eqnarray}
where the symbols $: ... :$ indicate that none of the quarks are contracted.  

Expanding the operator product into local operators we obtain a tower of 
four quark operators, which at tree level can be explicitly constructed
\begin{eqnarray} \label{Exp}
	&&  : (\bar{c}_L \gamma_\mu q_L)(x)  \,  (\bar{q}_L \gamma_\nu c_L)(0) : 
	=  : \bar{c}_L  \gamma_\mu q_L   \,  \bar{q}_L  \gamma_\nu c_L  :  
	\nonumber
	\\ 
	\nonumber 
	&& \quad \quad \quad \quad  
	+ x_\alpha :[\partial^\alpha  \bar{c}_L  \gamma_\mu q_L ]  \,  \bar{q}_L  \gamma_\nu c_L  :  
	\\ 
	&& \quad \quad \quad \quad 
	+ \frac{1}{2} x_\alpha x_\beta 
	:[\partial^\alpha \partial^\beta  \bar{c}_L  \gamma_\mu q_L ]  \,  \bar{q}_L  \gamma_\nu c_L  : 
	+ \cdots\,.
\end{eqnarray}
This expansion generates all tree-level contributions which show a phase-space enhancement by a factor 
of $16 \pi^2$ relative to the leading term. 
However, one could also use Eq.~(\ref{4q}) with an input from lattice QCD for the non-local 
matrix element. This corresponds to a partial resummation of (tree-level) contributions 
of the four quark operators generated by the above expansion. 

A commonly used ansatz for the evaluation of the non-local matrix element in Eq.~(\ref{Exp}) is VSA, 
which is defined through the replacement
\begin{eqnarray} \label{VSAAnn} 
	&& \langle D (v) |   : (\bar{c}_L \gamma_\mu q_L)(x)  \,  (\bar{q}_L \gamma_\nu c_L)(0) :   | D(v) \rangle 
	\\ 
	\nonumber 
	&& 
	\quad \sim  
	\langle D (v) |  (\bar{c}_L \gamma_\mu q_L)(x) | 0 \rangle \, \langle 0 | (\bar{q}_L \gamma_\nu c_L)(0)   | D(v) \rangle 
	\nonumber
	\\
	&& \quad\quad\quad
	= \frac{1}{4}f_D^2 M_D^2 v_\mu v_\nu \exp(i M_D (vx)) \,.
\end{eqnarray}  
Inserting this result into Eq.~(\ref{4q}) shows that all these contributions vanish in VSA. 
Since phenomenology tells us that VSA is a reasonable ansatz, the HQE for semileptonic 
decays of the ground state $D$ mesons  is not spoiled by 
the phase-space enhanced terms. 

The situation is different and a bit more involved in the case of nonleptonic $D$-meson decays. 
Keeping for simplicity only the Cabibbo allowed term we have 
\begin{equation} 
	H_{\rm eff,nl}  =   \frac{4 G_F}{\sqrt{2}}   V_{cs} V_{ud}^* 
	\left[C_1 (\mu) R_1 +  C_2 (\mu) R_2  \right] \,,
\end{equation}   
with 
\begin{equation} 
	R_1 = (\bar{c}_L^i \gamma_\mu  s_L^i) (\bar{d}_L^j \gamma^\mu u_L^j)\,, \quad 
	R_2 = (\bar{c}_L^i \gamma_\mu s_L^j) (\bar{d}_L^j \gamma^\mu u_L^i) \,,
\end{equation} 
where $i,j = 1,2,3$ are color indices. The Wilson coefficients 
depend on the scale $\mu$. 

In order to extract the four quark contributions at tree level it is
required to contract two of the quarks, which yields in total three contributions
corresponding to four quark operators with the quark content 
$(\bar{c}u)(\bar{u}c)$, $(\bar{c}s)(\bar{s}c)$ and $(\bar{c} d)(\bar{d} c)$. Since only 
two quarks are contracted, the Wilson coefficients of these contributions are phase-space enhanced 
compared to any other contribution. 

The contributions involving $(\bar{c}u)(\bar{u}c)$, $(\bar{c}s)(\bar{s}c)$ are usually called 
Weak Annihilation (WA) and can be treated in the same way
as the semileptonic case, the resulting expression reads 
\begin{eqnarray}
	&&\Gamma^{nl}_{\mbox{\scriptsize WA}} = \frac{G_F^2}{M_D} | V_{cs} V_{ud}^* |^2 
	\int \frac{d^4q}{(2\pi)^4} \frac{1}{3 \pi} (q^\mu q^\nu - g^{\mu \nu} q^2)   
	\label{WA}
	\\  
	&&  
	\quad
	\times \int d^4 x \, e^{-iqx}\bigg\{     
	 \bigg(3C_1^2 + 2C_1 C_2 + \frac{1}{3}C_2^2\bigg)
	\nonumber
	\\
	&&
	\quad \quad\quad
	\times  \langle D | : (\bar{c}_L \gamma_\mu s_L)(x)  \,  (\bar{s}_L \gamma_\nu  c_L)(0) : |D \rangle
	\nonumber
	\\
	&&
	\quad \quad \quad \quad \quad \quad \quad 
	+ \bigg(\frac{1}{3}C_1^2 + 2C_1 C_2 + 3C_2^2 \bigg)
	\nonumber
	\\
	&&
	\quad \quad\quad
	\times \langle D | : (\bar{c}_L \gamma_\mu u_L)(x)  \,  (\bar{u}_L \gamma_\nu  c_L)(0) : |D \rangle 
	\nonumber
	\\ 
	&& 
	\quad 
	+ 2C_2^2   \langle D | : (\bar{c}_L \gamma_\mu T^a s_L)(x)  \,  (\bar{s}_L \gamma_\nu T^a c_L)(0) : |D \rangle
	\nonumber
	\\ 
	&& 
	\quad 
	+ 2 C_1^2 \langle D | : (\bar{c}_L \gamma_\mu T^a u_L)(x)  \, (\bar{u}_L \gamma_\nu T^a c_L)(0) : |D \rangle    
	\bigg\} \,,
	\nonumber
\end{eqnarray}
while contribution involving $(\bar{c} d)(\bar{d} c)$ corresponds to Pauli interference (PI)  
\begin{eqnarray}
	&&\Gamma_{\mbox{\scriptsize PI}}^{nl} = \frac{G_F^2}{M_D} | V_{cs} V_{ud}^* |^2 
	\int \frac{d^4q}{(2\pi)^4} \frac{1}{3 \pi}   q^2 g^{\mu \nu}   \int d^4 x \, e^{-iqx}
	\nonumber
	\\ 
	&& 
	\quad \times \bigg\{ 
	6(C_1^2 + C_2^2)
	\nonumber
	\\
	&&
	\quad \quad \quad \times \langle D | : \bar{c}_L(x) \gamma_\mu T^a d_L(0)  \,  \bar{d}_L(x) \gamma_\nu T^a c_L(0) : |D \rangle 
	\nonumber 
        \\ 
	&& 
	\quad + (C_1^2 + 6C_1 C_2 + C_2^2)
	\nonumber
	\\
	&&
	\quad \quad \quad\times\langle D | : \bar{c}_L(x) \gamma_\mu d_L(0)  \,  \bar{d}_L(x) \gamma_\nu  c_L(0) : |D \rangle 
	 \bigg\} \,.
         \label{PI}
\end{eqnarray}
Expanding the $x$ dependence in the remaining field operators along the lines of Eq.~(\ref{Exp}) 
yields the tower of operators which have phase-space enhanced Wilson coefficients. To this end, 
the standard HQE takes the form 
\begin{equation} \label{enhance}
	\Gamma = \Gamma_0 + \sum_{k=1}^\infty a_k \left(\frac{\Lambda_{\rm QCD}}{m_Q} \right)^k 
	+ 16 \pi^2  \sum_{l=1}^\infty b_l  \left(\frac{\Lambda_{\rm QCD}}{m_Q} \right)^{l+3} \,, 
\end{equation}
where the second part originates from the (non-local) four quark 
operators for which the OPE generates a tower of local four quark operators.  
The coefficients $a_k$ and $b_l$ in Eq.~(\ref{enhance}) are  assumed to be of order 
unity. 

Inserting numerical values for the parameters we find for the charmed case 
(inserting $m_c = 1.5$ GeV and $\Lambda_{\rm QCD} = 500$ MeV)    
\begin{equation}
        \label{counting}
	16 \pi^2 \left(\frac{\Lambda_{\rm QCD}}{m_c} \right)^3  \sim 5 \,,
\end{equation} 
which clearly shows the need to re-organize the HQE for the case of charm, since the phase-space 
enhanced terms are as large as the leading piece. In fact, if 
we consider only the leading dimension-six term, we find a large negative contribution 
which overwhelms the piece from the free quark decay, leading overall to a negative 
for these two terms. Thus we need to keep the non-local expressions in 
Eqs.~(\ref{WA}) and (\ref{PI}), corresponding to a re-summation of the full tower
of four quark operators. 

Thus, for the charm quark we suggest to rearrange the HQE in Eq.~(\ref{enhance}) into a cHQE by 
\begin{equation} \label{NewHQE}
	\Gamma    
	= \Gamma_0^\prime + \sum_{k=1}^\infty a_k
	\left(\frac{\Lambda_{\rm QCD}}{m_Q} \right)^k  \,,
\end{equation}
with 
\begin{equation}
\Gamma_0^\prime = \Gamma_0 + 16 \pi^2  \sum_{l=1}^\infty b_l  \left(\frac{\Lambda_{\rm QCD}}{m_Q} \right)^{l+3} \,,
\end{equation}
where the second term at tree level within VSA is just Eq.\ (\ref{PI}). The key point of the cHQE is that 
all phase space enhanced terms are included into the leading term, while the remaining expansion
in Eq.~(\ref{NewHQE}) has coefficients $a_k \sim {\cal O}(1)$. 

For a quantitative discussion of the leading term one needs an input for these non-local matrix 
elements, which eventually will come from lattice QCD. However, one may as well just use a simple 
VSA based model for this, since the leading effect of this additional term is 
to fix the lifetime difference, and thus one could simply fit it to data.  

For illustration we inject such a simple model, defined by the replacement
\begin{equation} 
c(x) = e^{-i m_c (vx)} c(0)\,, \quad q(x) = e^{i \bar\Lambda (vx)} q(0) \,,
\end{equation}
with $\bar\Lambda = M_D - m_c$, which yields assuming VSA
\begin{eqnarray}
 &&\langle D^+ (v) | : \bar{c}_L(x) \gamma_\mu d_L(0)  \,  \bar{d}_L(x) \gamma^\mu  c_L(0) : |D^+ (v) \rangle 
 \nonumber
 \\ 
 &&\quad
 = \frac{1}{4} f_D^2 M_D^2 \exp[i (m_c - \bar\Lambda) vx ] \,,
 \label{modelPIME}
\end{eqnarray}
while all other matrix elements (in particular all matrix elements involving neutral $D$ mesons)
are assumed to vanish. Evaluating this yields for the leading term 
\begin{eqnarray} 
\Gamma_0^\prime &=& \frac{G_F^2 m_c^5 |V_{cs} V_{ud}^*|^2}{192 \pi^3} \\ \nonumber 
&& \times \left[ \kappa + \kappa'  16\pi^2  \bigg(\frac{f_D^2 M_D}{m_c^3}\bigg)\bigg(1- \frac{\bar\Lambda}{m_c}\bigg)^2\right],
\end{eqnarray}
with $\kappa = 3C_1^2+3C_2^2+2C_1C_2$ and $\kappa^\prime = C_1^2+C_2^2+6C_1C_2$. 

In fact, the leading term of cHQE reproduces the nonleptonic lifetime ratio of the charged and
neutral $D$ meson. Using the numerical values for the Wilson coefficients 
at the relevant scale $\mu \sim m_c$, namely $C_1 (m_c) = 1.25$ and $C_2 (m_c) = -0.49$
\cite{Buchalla:1995vs}, we obtain $\kappa = 4.18$ and  $\kappa' = -1.89$. Inserting 
$m_c = 1.36\mbox{ GeV}$ and $\bar{\Lambda} = 0.61\mbox{ GeV}$
we can compute the nonleptonic width ratio
\begin{equation}
 N^{(th)}(D^\pm,D^0) = \frac{\Gamma^{nl}(D^\pm)}{\Gamma^{nl}(D^0)} = 0.3\,,
\end{equation}  
which is to be compared to the experimental number
\begin{equation}
 N^{(exp)}(D^\pm,D^0) = 0.301 \pm 0.005\,. 
\end{equation}
This shows that the corrections to this leading term will be small.  

These corrections can be computed in an expansion in 
powers of $\Lambda_{\rm QCD}/m_c$ with coefficients of order unity. 
In addition, we have the usual perturbative expansion of the coefficients in powers of 
$\alpha_s (m_c)$, which also holds for the phase space enhanced terms. It is a matter of
taste, if one includes these into $\Gamma_0^\prime$ or keeps them as small terms in the expansion. 

The problem that neither $\Lambda_{\rm QCD}/m_c$ nor $\alpha_s (m_c)$ is particularly small 
is not solved, rather we only tame the large ${\cal O}(1)$ terms in the HQE by switching to 
cHQE. The remaining problems manifest themselves in large uncertainties induced by the 
strong dependence on the charm mass for some of the observables, since the size of the QCD 
corrections depends on the choice of the mass scheme. To what extend the cHQE can be turned 
into a precision tool, similar to what we have in the bottom sector, remains to be explored.

\subsection*{Acknowledgments}

We are grateful to Alexander Lenz for discussions
on the HQE and its applications to charm physics,
and Maria Laura Piscopo for valuable communications. 
This research was supported by the Deutsche Forschungsgemeinschaft 
(DFG, German Research Foundation) under grant  396021762 - TRR 257 
``Particle Physics Phenomenology after the Higgs Discovery''.

%\appendix

%\section{Appendixes}


\begin{thebibliography}{99}

% citations HQET 	


%\cite{Shifman:1987rj}
\bibitem{Shifman:1987rj}
M.~A.~Shifman and M.~B.~Voloshin,
%``On Production of d and D* Mesons in B Meson Decays,''
Sov. J. Nucl. Phys. \textbf{47} (1988), 511
ITEP-87-64.
%670 citations counted in INSPIRE as of 09 Feb 2021

%\cite{Eichten:1989zv}
\bibitem{Eichten:1989zv}
E.~Eichten and B.~R.~Hill,
%``An Effective Field Theory for the Calculation of Matrix Elements Involving Heavy Quarks,''
Phys. Lett. B \textbf{234} (1990), 511-516
doi:10.1016/0370-2693(90)92049-O
%1097 citations counted in INSPIRE as of 09 Feb 2021

%\cite{Isgur:1989vq}
\bibitem{Isgur:1989vq}
N.~Isgur and M.~B.~Wise,
%``Weak Decays of Heavy Mesons in the Static Quark Approximation,''
Phys. Lett. B \textbf{232} (1989), 113-117
doi:10.1016/0370-2693(89)90566-2
%2302 citations counted in INSPIRE as of 09 Feb 2021

%\cite{Grinstein:1990mj}
\bibitem{Grinstein:1990mj}
B.~Grinstein,
%``The Static Quark Effective Theory,''
Nucl. Phys. B \textbf{339} (1990), 253-268
doi:10.1016/0550-3213(90)90349-I
%570 citations counted in INSPIRE as of 09 Feb 2021 

%\cite{Manohar:2000dt}
\bibitem{Manohar:2000dt}
A.~V.~Manohar and M.~B.~Wise,
%``Heavy quark physics,''
Camb. Monogr. Part. Phys. Nucl. Phys. Cosmol. \textbf{10} (2000), 1-191
%499 citations counted in INSPIRE as of 09 Feb 2021

% citations HQE 

%\cite{Chay:1990da}
\bibitem{Chay:1990da}
J.~Chay, H.~Georgi and B.~Grinstein,
%``Lepton energy distributions in heavy meson decays from QCD,''
Phys. Lett. B \textbf{247} (1990), 399-405
doi:10.1016/0370-2693(90)90916-T
%623 citations counted in INSPIRE as of 09 Feb 2021

%\cite{Bigi:1992su}
\bibitem{Bigi:1992su}
I.~I.~Y.~Bigi, N.~G.~Uraltsev and A.~I.~Vainshtein,
%``Nonperturbative corrections to inclusive beauty and charm decays: QCD versus phenomenological models,''
Phys. Lett. B \textbf{293} (1992), 430-436
[erratum: Phys. Lett. B \textbf{297} (1992), 477-477]
doi:10.1016/0370-2693(92)90908-M
[arXiv:hep-ph/9207214 [hep-ph]].
%657 citations counted in INSPIRE as of 09 Feb 2021

%\cite{Manohar:1993qn}
\bibitem{Manohar:1993qn}
A.~V.~Manohar and M.~B.~Wise,
%``Inclusive semileptonic B and polarized Lambda(b) decays from QCD,''
Phys. Rev. D \textbf{49} (1994), 1310-1329
doi:10.1103/PhysRevD.49.1310
[arXiv:hep-ph/9308246 [hep-ph]].
%612 citations counted in INSPIRE as of 09 Feb 2021

%\cite{Benson:2003kp}
\bibitem{Benson:2003kp}
D.~Benson, I.~I.~Bigi, T.~Mannel and N.~Uraltsev,
%``Imprecated, yet impeccable: On the theoretical evaluation of Gamma(B ---\ensuremath{>} X(c) l nu),''
Nucl. Phys. B \textbf{665} (2003), 367-401
doi:10.1016/S0550-3213(03)00452-8
[arXiv:hep-ph/0302262 [hep-ph]].
%175 citations counted in INSPIRE as of 09 Feb 2021

%\cite{Gambino:2013rza}
\bibitem{Gambino:2013rza}
P.~Gambino and C.~Schwanda,
%``Inclusive semileptonic fits, heavy quark masses, and $V_{cb}$,''
Phys. Rev. D \textbf{89} (2014) no.1, 014022
doi:10.1103/PhysRevD.89.014022
[arXiv:1307.4551 [hep-ph]].
%89 citations counted in INSPIRE as of 09 Feb 2021

%\cite{Zyla:2020zbs}
\bibitem{Zyla:2020zbs}
P.~A.~Zyla \textit{et al.} [Particle Data Group],
%``Review of Particle Physics,''
PTEP \textbf{2020} (2020) no.8, 083C01
doi:10.1093/ptep/ptaa104
%750 citations counted in INSPIRE as of 09 Feb 2021

\bibitem{Guberina:1979xw}
B.~Guberina, S.~Nussinov, R.~D.~Peccei and R.~Ruckl,
%``D Meson Lifetimes and Decays,''
Phys. Lett. B \textbf{89} (1979), 111-115.
%doi:10.1016/0370-2693(79)90086-8
%258 citations counted in INSPIRE as of 03 Feb 2021

\bibitem{Neubert:1996we}
M.~Neubert and C.~T.~Sachrajda,
%``Spectator effects in inclusive decays of beauty hadrons,''
Nucl. Phys. B \textbf{483} (1997), 339-370.
%doi:10.1016/S0550-3213(96)00559-7
%[arXiv:hep-ph/9603202 [hep-ph]].
%329 citations counted in INSPIRE as of 15 Apr 2020

\bibitem{Lenz:2013aua}
A.~Lenz and T.~Rauh,
%``D-meson lifetimes within the heavy quark expansion,''
Phys. Rev. D \textbf{88} (2013), 034004.
%doi:10.1103/PhysRevD.88.034004
%[arXiv:1305.3588 [hep-ph]].

%\cite{Bauer:1986bm}
\bibitem{Bauer:1986bm}
M.~Bauer, B.~Stech and M.~Wirbel,
%``Exclusive Nonleptonic Decays of D, D(s), and B Mesons,''
Z. Phys. C \textbf{34} (1987), 103
doi:10.1007/BF01561122
%1859 citations counted in INSPIRE as of 09 Feb 2021

%\cite{Buchalla:1995vs}
\bibitem{Buchalla:1995vs}
G.~Buchalla, A.~J.~Buras and M.~E.~Lautenbacher,
%``Weak decays beyond leading logarithms,''
Rev. Mod. Phys. \textbf{68} (1996), 1125-1144
doi:10.1103/RevModPhys.68.1125
[arXiv:hep-ph/9512380 [hep-ph]].
%2629 citations counted in INSPIRE as of 09 Feb 2021

\end{thebibliography}
\end{document}